\documentclass[prl,twocolumn,showpacs]{revtex4}

\usepackage{amsmath}
\usepackage{amssymb}
\usepackage{graphicx}
\usepackage{mathrsfs}

\begin{document}

\title{Self-compression and catastrophic collapse of photon bullets in
  vacuum} 


\author{Mattias Marklund}
\email{marklund@elmagn.chalmers.se}
\altaffiliation[Also at: ]{Department of Electromagnetics, Chalmers
  University of Technology, SE--412 96 G\"oteborg, Sweden}
\affiliation{Fakult\"at f\"ur Physik und Astronomie, 
  Ruhr-Universit\"at Bochum, D--44780 Bochum, Germany}

\author{Bengt Eliasson} 
\affiliation{Fakult\"at f\"ur Physik und Astronomie, 
  Ruhr-Universit\"at Bochum, D--44780 Bochum, Germany}

\author{Padma K.\ Shukla}
\altaffiliation[Also at: ]{Department of Physics, Ume{\aa} University,
  SE--901 87 Ume{\aa}, Sweden}
\affiliation{Fakult\"at f\"ur Physik und Astronomie, 
  Ruhr-Universit\"at Bochum, D--44780 Bochum, Germany}

\date{January 5, 2004}

\begin{abstract}
Photon--photon scattering, due to photons interacting with virtual
electron--positron pairs, is an intriguing deviation from classical
electromagnetism predicted by quantum electrodynamics (QED). Apart
from being of fundamental interest in itself, collisions between
photons are believed to be of importance in the vicinity of magnetars,
in the present generation intense lasers, and in intense
laser-plasma/matter interactions; the latter recreating astrophysical
conditions in the laboratory. We show that an intense photon pulse
propagating through a radiation gas can self-focus, and under certain
circumstances collapse. This is due to the response of the radiation
background, creating a potential well in which the pulse gets trapped,
giving rise to photonic solitary structures. When the radiation gas
intensity has reached its peak values, the gas releases part of its
energy into `photon wedges', similar to Cherenkov radiation. The
results should be of importance for the present generation of intense
lasers and for the understanding of localized gamma ray bursts in
astrophysical environments. They could furthermore test the
predictions of QED, and give means to create ultra-intense photonic
pulses.  
\end{abstract} 

\maketitle

In classical electrodynamics, as described by the Maxwell equations,
photons do not interact as long as there is no material medium present.  
However, due to the interaction of photons with virtual electron--positron 
pairs, quantum electrodynamics (QED) predicts photon--photon scattering in 
vacuum \cite{Schwinger}. This is commonly modeled by the Heisenberg--Euler 
(H-E) Lagrangian, which neglects dispersive effects. The H-E Lagrangian gives 
rise to cubic nonlinear corrections to Maxwell's vacuum equations, similar 
to the self-interaction terms encountered in optics of Kerr media
\cite{Kivshar-Agrawal}.  
The H-E corrections give rise  to both single particle effects, such
as photon splitting 
\cite{Adler,Baring-Harding},  lensing effects in strong magnetic
fields \cite{Harding},  
like the ones in magnetar environments \cite{Kouveliotou}, and to
coherent field  
effects such as harmonic generation \cite{Ding-Kaplan} and
self-focusing of photon 
beams \cite{Rozanov}. Efforts to detect these collisions are being
made by using  
state-of-the-art super-conducting microwave facilities
\cite{Brodin-Marklund-Stenflo}.    
Recently, it has been shown theoretically that QED effects can give
rise to two-dimensional   
collapsing photonic structures in a radiation gas
\cite{Marklund-Brodin-Stenflo}, which 
could be of importance for photon propagation in stellar atmospheres
and in the early  
Universe. Studies of photon pulses in a radiation background and of
optical pulses in  
nonlinear media reveal that they share common features which can be described
mathematically by a Schr\"odinger equation with a nonlinear potential
\cite{Rozanov,Brodin-Marklund-Stenflo}.  

Dispersive effects can play an important role for short optical pulses
when the spatial gradients and time variations become large. One of the 
most important generic effects of the dispersion is to permit pulse splitting 
along the direction of propagation \cite{Chernev-Petrov,Rothenberg}, which has 
also been experimentally verified \cite{Ranka-Schirmer-Gaeta}. The
pulse splitting  
is of great interest in applications to normal dispersive media, where
the collapse  
of light pulses can be arrested (which is not possible in anomalous
dispersive media 
\cite{Silberberg}), giving rise to a train of high-intensity pulses;
these pulses  
may then work as a source of white light generation \cite{Gaeta2}. The quantum 
vacuum also posses such dispersive effects \cite{Rozanov}, which can
be of importance   
in the present generation intense lasers \cite{Mourou-Barty-Perry}
where ultra-short  
high-intensity photon pulses will be produced. Rapidly varying fields
are  also of  
interest for the frequency up-shift of photons in photon acceleration 
\cite{photonacceleration,Mendonca1,Mendonca3}, which is an important
ingredient in  
studies of plasma-based charged particle accelerators and laser-plasma/matter 
interactions \cite{Pukhov} where ultra-strong fields \cite{Malka-etal,   
Pukhov-Gordienko-Baeva,Bingham} are expected to reach from peta- to zeta-watt 
powers \cite{Bulanov-Esirkepov-Tajima}.  In this Letter, we present
for the first time results dealing with the nonlinear propagation of 
{\it three-dimensional intense photon pulses} in vacuum where QED effects
play a major role. Specifically, we report on new features of
nonlinear propagation 
of a linearly polarized intense photon pulse on a radiation gas
background, assuming 
that there is no pair creation and that the field strength $E$ is
below the critical 
Schwinger field, i.e. $ \omega \ll m_ec^2/\hbar \simeq 8\times
10^{20}\,\text{rad}\,\mathrm{s}^{-1}$ and $|E| \ll  
m_ec^2/e\lambda_c \simeq 10^{18}\,\mathrm{V}\mathrm{m}^{-1}$,
respectively, where $\omega$ is the photon frequency, $m_e$ the electron mass, 
$e$ the magnitude of the electron charge, and $\lambda_c$ the Compton
wavelength.  
The derivative corrections to the H-E Lagrangian gives rise to a
nonlinear dependence  
of the photon frequency on the wavenumber, which is shown to permit
three-dimensional  
self-focusing. For moderate intensities of the photon pulse, the
self-focusing will be  
followed by pulse splitting and the formation of stable photonic
solitary-pulses.  
However, for high initial powers of the photon pulse, the latter
undergoes catastrophic  
collapse, giving rise to field amplitudes exceeding the Schwinger limit 
$10^{29}\,\mathrm{W\,cm^{-2}}$.  At these intensities, our theory
breaks down and  
one must consider higher-order nonlinear effects and pair
productions. It is argued  
that higher-order nonlinear effects become important before the
Schwinger limit is  
reached, arresting the collapse and giving rise to ultra-high
intensity three-dimensional  
solitary photonic pulses.  Thus, the results presented here gives,
apart from its 
immediate fundamental interest and astrophysical applications, a
mechanism for creating  
ultra-high intensity photon pulses.  
 
The evolution of an intense short photon pulse and the radiation
background, is  
governed by the Karpman-like system of equations
\cite{Marklund-Brodin-Stenflo,Rozanov,Karpman1,Karpman2}   
\begin{subequations}
\begin{equation}
  i\left(\frac{\partial}{\partial t} + v_g\frac{\partial}{\partial z}
  \right)E + \frac{v_g}{2k_0}\left[
  \frac{1}{r}\frac{\partial}{\partial r}\left(r\frac{\partial
  E}{\partial r}\right) - 
    \beta_{z}\frac{\partial^2 E}{\partial z^2} \right] +
  \kappa\mathscr{E}E = 0 ,  
\label{eq:nlse}
\end{equation}
and
\begin{equation}
  \left( \frac{\partial^2}{\partial t^2} - \frac{c^2}{3}\nabla^2
  \right)\mathscr{E} + \mu\epsilon_0\left(
  \frac{\partial^2}{\partial t^2} + {c^2}\nabla^2 
  \right) |E|^2 = 0,
\label{eq:response}
\end{equation}
\label{eq:system}
\end{subequations}
where $\mathscr{E}_0$ and $\mathscr{E}$ is the background and
perturbation of the  
radiation energy density, respectively, $v_g = c(1 - \mu -
\mu^2\delta)$ is the group  
velocity, $\beta_{z} \simeq 2\mu^2\delta$ the vacuum dispersion
coefficient, and 
$\kappa \simeq ck_0(1 + \mu\delta)/\mathscr{E}_c$ the vacuum nonlinear
refraction parameter \cite{Marklund-Brodin-Stenflo}. 
Here, $\delta \equiv k_0^2/k_c^2$ and $\mu \equiv
\mathscr{E}_0/\mathscr{E}_c$, and the critical parameters $k_c^{-1} \sim 
10^{-13}\,\mathrm{m}$, the Compton wavelength divided by $2\pi$,  and
$\mathscr{E}_c \sim 10^{27}\,\mathrm{Jm^{-3}}$ are defined by the QED
properties of the vacuum.  

%
%


If the time response of the radiation background is slow, then Eq.\
(\ref{eq:response}) may be integrated to yield 
$\mathscr{E}
\simeq 3\mu\epsilon_0|E|^2$ and from Eq.\ (\ref{eq:nlse}) we obtain the  
standard equation for analyzing ultra-short intense pulses in normal
dispersive media, see Refs.\
\cite{Chernev-Petrov,Rothenberg,Kivshar-Agrawal}   
and references therein. It is well-known that the evolution of a  pulse 
within this equation is modulationally unstable \cite{Kivshar-Agrawal},
and displays first self-focusing, then pulse splitting along the direction 
of the pulse propagation \cite{Chernev-Petrov,Rothenberg}.     

\begin{figure}[ht]
\centerline{\includegraphics[width=\columnwidth]{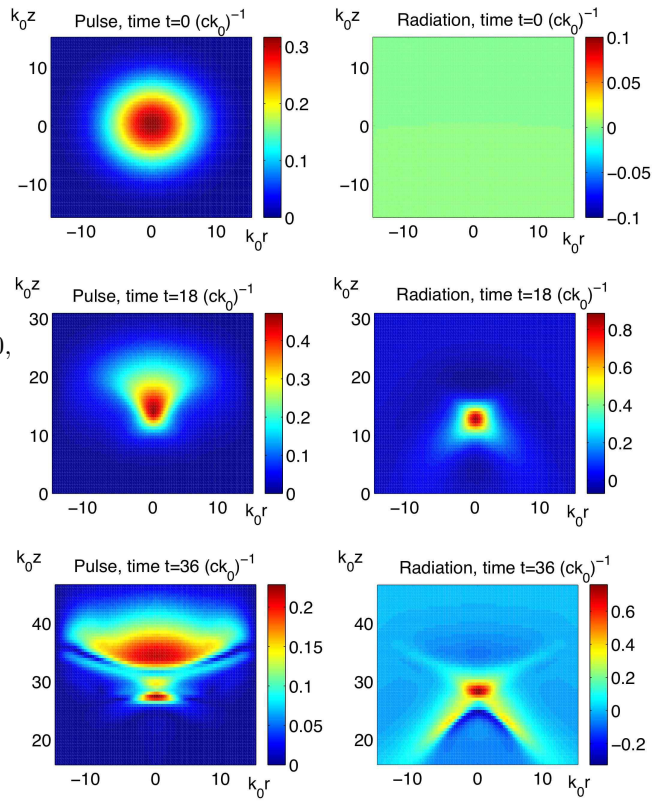}}
\caption{The collapse of the pulse and radiation gas disturbance
  displayed using normalized electric field intensity $|E|^2$ (left
  panels)  and energy densities $\mathscr{E}$ (right panels). We have 
  scaled the pulse electric field $E$ by
  $2(3\sqrt{\mu\epsilon_0/\mathscr{E}_c})^{-1}$   
  and the radiation background $\mathscr{E}$ by 
  $2\mathscr{E}_c/3$, while the time scale is normalized
  by $(ck_0)^{-1}$ and the length scale by $k_0^{-1}$. The initially
  Gaussian pulse has the amplitude
  $E_0=0.64(3\sqrt{\mu\epsilon_0/\mathscr{E}_c})^{-1}$ and width 
  $a_0^2 = 50/k_0^2$. The upper panels show the initial configuration, while
  the middle panels give the system after 18 time units and the lower
  panels after 36 time units. The initial photonic pulse in the upper
  left panel self-focuses and then splits into two pulses, as can be
  seen in the left middle and lower panels. The energy background
  emanates Cherenkov-like radiation, clearly seen in the lower right
  panel as two yellow bands created behind the pulse.} 
\end{figure}


We have analyzed the system (\ref{eq:system}) numerically and
analytically; in Fig.~1, we display the intensity of
the electric field $E$ in the left panels and the perturbation
$\mathscr{E}$ of the background radiation in the right panels. As an
initial condition, we use a Gaussian pulse for the electric field
envelope, $E=E_0\exp\{-[r^2 + (z + z_0)^2]/a_0^2\}$,
while the radiation perturbation is initially set to zero, and $v_g \simeq
c$ together with $\beta_z = 10^{-7}$ (relevant to astrophysical
applications); see the upper panels of Fig.~1. The $z$
derivatives in 
Eq.~(\ref{eq:system}) are calculated numerically with a pseudo-spectral
method, the $r$ derivatives with a second-order 
difference scheme, and the system is advanced in time with a
fourth-order Runge--Kutta scheme.

Displayed in the middle panels of Fig.~1 is 
the initial pulse which has been moving in the $z$ direction with 
a speed close to the speed of light. The pulse has self-focused, 
and exhibits a structure stretched along the $z$ direction. The 
self-focusing can be understood in the framework of the two-dimensional 
nonlinear Schr\"odinger equation; during the initial phase of the pulse
propagation, the dispersive term plays a minor dynamical role. 
If the $t$ and $z$ variations are neglected in Eq.\ (\ref{eq:response}), 
we obtain $\mathscr{E} \simeq 3\mu\epsilon_0|E|^2$. 
Thus, Eq.\ (\ref{eq:response}) becomes the standard, 
nonlinear Schr\"odinger equation in which the cubic nonlinear 
potential has the same sign as the diffraction term, supporting
two-dimensional collapse of the pulse \cite{Kivshar-Agrawal}. The
time for \emph{complete} two-dimensional collapse of a pulse can be
estimated by means of Rayleigh--Ritz optimization. Using normalized units 
(see Fig.\ 1), two-dimensional collapse will occur when the intensity and 
pulse width satisfies the inequality $|E_0|^2 \gtrsim 1.3 a_0^{-2}$.  
The resulting collapse time $t_c \simeq 4.3 a_0^2 ( E_0^2a_0^2 -
1.3)^{-1/2}$. In Fig.\ 1, pulse splitting occurs after $\sim t_c/3$,
thus well before field strengths reaches critical levels. 

However, here the collapse of the pulse is arrested due to the backscattering 
from the background; the core of the pulse slows down because of the nonlinear 
interaction with the background, while the flanks of the pulse continue to 
propagate with the speed of light, creating a fan-like structure in front of 
the slower moving pulse; see the middle left panel of Fig.~1. 
The pulse then splits into several parts with local maxima; one wider
pulse which  
can be seen at $k_0z=35$ in the lower left panel followed by two
smaller, narrower  
pulses at $k_0z\approx27$.  We have carried out simulations with
different values  
of $\beta_z$, which reveal that the flattening of the leading edge
pulse is less 
pronounced for smaller values of $\beta_z$. The wider pulse is
correlated with a  
slight depletion of the radiation energy density. Letting the pulse
propagate further,  
the transverse variation for the wide pulse will become much smaller
than the longitudinal  
variation, leading to an almost one-dimensional structure (depending
weakly on the 
transverse coordinate $r$), thus making the nonzero dispersive term
essential.   
This pulse moves with a supersonic speed ($>c/\sqrt{3}$). For
these broad supersonic  
pulses, we may solve Eqs.\ (\ref{eq:system}) for the stationary
state. The acoustic 
equation (\ref{eq:response}) gives  
$\mathscr{E} = -3\mu(v_g^2 + c^2)/(3v_g^2 - c^2)\epsilon_0|E|^2$. We
note that the pulse will experience resonance phenomena as the group
velocity approaches the sound speed. 
With the solution to Eq.\ (\ref{eq:response}), one can describe the
localized stationary solitary solutions for a pulse moving close to
the speed of light according to \cite{Kivshar-Agrawal}  
$|E| \simeq E_0\,\text{sech}\,[(3\varepsilon/\delta)^{1/2}k_0(z - v_g
  t)]$, where $E_0$ is the constant amplitude of the pulse, and
$\varepsilon = \epsilon_0E_0^2/\mathscr{E}_0$ is the relative energy
density of the pulse. 
We point out that these pulses are not necessarily of extremely
high amplitudes. The remnants of radially collapsing 
pulse asymptotically form a train of low-amplitude axially modulated
solitary pulses. Moreover, the pulses show a parabolic self-compression, 
which will compensate for the small, but non-zero, diffraction of the pulse, 
making the one-dimensional approximation valid over a longer period of time. 
%
In the final stage, we can see the formation of a `photon wedge'
(seen in the lower right panel,) which is due to the pulse
propagation with a supersonic speed, so that part of the energy
of the pulse is released into Cherenkov-like radiation behind the pulse,
analogous to the sonic shocks created behind supersonic flights in air.
Thus, the dynamics of the photonic pulse is surprisingly complex, 
exhibiting a multitude of nonlinear phenomena.

Simulations with higher initial amplitudes of the pulse show that the pulse 
again experience two-dimensional collapse in which the  intensity may grow 
above the Schwinger limit before photon pulse splitting occurs. Because of the 
nonlinear dominance,  the speed of the pulse decreases below the sound
speed of  
the background, and the Cherenkov energy loss through the photon wedges will 
therefore be small, reinforcing the photon pulse collapse. This
mechanism leads to 
amplitudes where our model breaks down; thus, the photon collapse
becomes catastrophic   
close to the Schwinger limit \cite{Gaeta2}. However, before the
Schwinger limit is  
reached, higher order nonlinear effects are likely to arrest the
collapse \cite{DT}.   
This effect is similar to ultra-short intense laser pulses in air,
where the formation  
of a plasma due to self-focusing gives rise to filamentation and
halted collapse 
\cite{Tzortzakis}. The plasma formation gives rise to a significantly longer
propagation range for laser pulses in air, a behavior which the QED pulse 
propagation presented here is expected to share, but with the plasma
formation replaced by higher order QED effects and the pair creation. 

Astrophysical environments can be of very extreme nature, exhibiting
the largest energy levels known to man. In the case of regular neutron
stars the surface magnetic field strengths reach $10^{10}-10^{13}\,\mathrm{G}$,
while in magnetars they can reach $10^{14}-10^{15}\,\mathrm{G}$, the
latter being close to the Schwinger limit. An interesting possibility
arises in the context of neutron star and magnetar quakes, in which magnetic
fields build up tensions in the star crust over long periods of time,
and sudden bursts of energy are released from the star during the quakes. 
There, it is expected that large quantities of low-frequency photons would be 
ejected, forming an almost incoherent spectrum of waves \cite{Kondratyev}. 
This photon gas could reach energy densities $\mathscr{E}_0 \sim
10^{17} - 10^{26} 
\,\mathrm{J\,m^{-3}}$, corresponding to $\mu \sim 10^{-10} - 10^{-1}$. A short 
high intensity electromagnetic pulse, with wavelengths from the UV to
gamma range 
(corresponding to $\delta \sim 10^{-10} - 10^{-2}$), with its
evolution modeled by  
the nonlinear Schr\"odinger equation (\ref{eq:nlse}), passing through this
low-frequency photon gas dynamically governed by the acoustic-like
wave equation (\ref{eq:response}), could be a source for gamma-ray
bursts.  The latter are short ($\lesssim$ tens of second) emissions of
photons in  
the gamma range \cite{Piran}. Apart from the pulse collapse described by 
the system (\ref{eq:system}),  there will also be a significant blue-shift 
of the photon pulse, due to the formation of steep intensity gradients
in the system \cite{Gaeta2}, which we can see in the numerical results
of Fig.~1.  We note that the timescales for these
events could be extremely short depending on the frequency of the
photons. 
One can therefore expect the formation of intense photon pulses with
frequencies  
up to the gamma regime, within the incoherent photon gas created by
magnetar quakes,  
possibly giving insight into the dynamics of gamma-ray bursts, since
we here have a 
mechanism both for blue shifting, pulse compression and high intensity
field generation. However, we note that, in accordance with the standard 
relativistic fireball model, it would be more appropriate to model incoherent 
photons by a wave kinetic equation, instead of a Schr\"odinger
equation. 

To summarize, we have considered the implications of the dispersive 
properties of the quantum vacuum for the case of intense photon pulses 
propagating on a radiation background. In the slowly varying acoustic wave 
limit, the pulse evolves similar to an ultra-short high-intensity pulse in 
nonlinear, normal dispersive media, with pulse collapse and splitting as a 
result. The analysis of the full system of equations shows that the slowly 
varying acoustic limit is far from generic, and that the response of the
radiation gas can have the same timescale as the pulse evolution. It
is due to the self-generation of potential wells, giving an attractive
force between the photonic pulse peak and the acoustic disturbance.  This
can give rise to three-dimensional catastrophic photonic pulse collapse, 
where the pulse and radiation gas power increases towards the
Schwinger limit.  Moreover, given suitable initial conditions, the photonic 
pulses can evolve into a stable localized structure with high field
strengths. The application of our work to astrophysical 
settings  has been discussed. Specifically, the present investigation sheds 
light on gamma-ray burst dynamics, and gives a means for obtaining pulse 
intensities surpassing the ones achievable by known mechanisms.    

\acknowledgments
This work was partially supported by the European Commission
(Brussels, Belgium)   
through contract No. HPRN-CT-2000-00314 for carrying out the task of the Human
Potential Research Training Networks ``Turbulent Boundary Layers in
Geospace Plasmas'',  as well as by the Deutsche Forschungsgemeinschaft 
(Bonn, Germany) through the Sonderforschungsbereich 591 entitled
``Universelles  
Verhalten Gleichgewichtsferner Plasmen: Heizung, Transport und
Strukturbildung''. 


\end{document}